# School value-added models for multivariate academic and non-academic outcomes: A more rounded approach to using student data to inform school accountability

Lucy Prior, Harvey Goldstein, George Leckie

Centre for Multilevel Modelling and School of Education, University of Bristol

**Corresponding author:**

lucy.prior@bristol.ac.uk, School of Education, 35 Berkeley square, Bristol, BS8 1JA

**Acknowledgements:**

This work was produced using data from the Department for Education National Pupil Database, distributed by the ONS which is Crown Copyright. The use of the ONS statistical data in this work does not imply the endorsement of the ONS in relation to the interpretation or analysis of the statistical data.

**Funding:**

This work was supported by the Economic and Social Research Council under Grant number ES/R010285/1.

# School value-added models for multivariate academic and non-academic outcomes: A more rounded approach to using student data to inform school accountability

**Abstract**

Education systems around the world increasingly rely on school value-added models to hold schools to account. These models typically focus on a limited number of academic outcomes, failing to recognise the broader range of non-academic student outcomes, attitudes and behaviours to which schools contribute. We explore how the traditional multilevel modelling approach to school value-added models can be extended to simultaneously analyse multiple academic and non-academic outcomes and thereby can potentially provide a more rounded approach to using student data to inform school accountability. We jointly model student attainment, absence and exclusion data for schools in England. We find different results across the three outcomes, in terms of the size and consistency of school effects, and the importance of adjusting for student and school characteristics. The results suggest the three outcomes are capturing fundamentally distinct aspects of school performance, recommending the consideration of non-academic outcomes in systems of school accountability.

Keywords: school accountability, value-added, multilevel model, absenteeism, exclusions

# 1. Introduction

The use of student attainment data and school value-added models for school accountability is becoming increasingly widespread among school systems around the world (OECD, 2008; NFER, 2018). Summaries of student performance data, from exams or standardised tests, are used to judge the effectiveness of schools and can determine funding decisions, whether schools are taken over by competing establishments or even closed. Furthermore, these summaries are often publicly disseminated in the form of high-profile school league tables (Leckie & Goldstein, 2009). The almost exclusive focus on student attainment in most accountability systems means they often fail to recognise the full breadth of student outcomes, attitudes and behaviours that schools influence. Additionally, the high-stakes nature of school accountability has led to criticisms regarding perverse incentives, 'gaming' behaviour, and many other deleterious consequences (Ehren & Swanborn, 2012; Foley & Goldstein, 2012; West, 2010). Basing school summaries of student performance on a wider range of outcomes, with the aspiration of making more holistic and sensitive judgements of schools, may provide a way to address these two important criticisms routinely levelled at many accountability systems. This article explores these issues by expanding the traditional school value-added model of student attainment to simultaneously encompass two important non-academic student outcomes, absences and exclusions, both of which have been shown to vary across schools and have notable relationships with academic performance (Garcia & Weiss, 2018; Gottfried, 2019; Sullivan, et al., 2013; Timpson, 2019).

School value-added models have become the de facto approach to estimating school effects on student attainment (Reynolds, et al., 2014). Rather than directly compare schools in terms of average student final attainment (which largely reflects school differences in attainment at intake), the value-added approach compares schools in terms of the average progress or change in attainment made by students in each school over the current phase of schooling. As such, school value-added models are argued to offer fairer comparisons between schools, with the minimum requirement being a measure of prior attainment (Goldstein, 1997; Leckie, et al., 2019) and arguments are often made for the additional control of student background characteristics which also vary between schools and are predictive of subsequent learning (Goldstein, et al., 2000; Leckie & Goldstein, 2019; Timmermans & Thomas, 2015).

However, school value-added models customarily focus on student academic outcomes leading to potentially narrow summaries of school quality. While student academic and non-academic outcomes are likely to be correlated there is no guarantee that schools which prove effective for the former will prove effective for the latter. Indeed, even when we restrict our attention to academic outcomes we see inconsistency in school effects across academic subjects such as English, maths, science and so on (Marks, 2015; Reynolds, et al., 2014; Teddlie & Reynolds, 2000; Thomas, et al., 1997). Even greater inconsistency in school effects might reasonably be expected were we to additionally focus on non-academic outcomes. This article explores how school value-added models can be extended to achieve this in order to support a more holistic approach to school accountability. One in which schools are recognised as playing an integral role in 'whole' child development and as vital links in chains of social exclusion or inclusion. Such a viewpoint has been codified in international and national policies such as the 2015 Every Student Succeeds Act (ESSA, 2015) in the USA which obliges states to report a non-academic performance measure, with most states choosing measures of absenteeism to fulfil this requirement (Garcia & Weiss, 2018), and the EU 2020 Strategy for which the number of early school leavers is an indicator (European Commission, 2010). We focus on student absences and exclusions as two important non-academic outcomes to be considered alongside the traditional focus on student attainment.

Low absence rates are a key student and school outcome; absence is associated with lower attainment (Garcia & Weiss, 2018; Hancock, et al., 2017). Additionally, higher rates of absence within schools may have knock-on effects for the class-peers of absentees: teachers must accommodate the catch-up of those who have missed lessons (Gottfried, 2019). Reciprocal relationships between school disengagement, alienation and absenteeism have also been hypothesised (Ramberg, et al., 2019; Rothman, 2001). Therefore, monitoring school rates of absence could prove a useful screening tool to identify schools where student-school relationships may have broken down. Chronic absenteeism has also been associated with negative later-life outcomes, such as alcohol and drug use, school drop-out and unemployment (Ingul, et al., 2012; Gottfried, 2009). Investigation and intervention as part of a system of school accountability that utilises student absence data may mitigate some of these detrimental outcomes. Absence measures are already reported in some educational systems, for instance England (Department for Education, 2019a) and Australia (https://www.myschool.edu.au/), however, this is often in the form of unadjusted rates. Consequently, there appears a need for a value-added type approach to adjust school absence rates for school intake differences in order to achieve fairer and more meaningful comparisons, as is now standard practice when comparing school average attainment scores.

Exclusion from school is also associated with lower academic performance and negative later-life outcomes (Timpson, 2019). For instance, students who are excluded are more likely to end up in prison or involved in criminal justice systems (Gill, et al., 2017; Sanders, et al., 2018). Variation between schools in exclusion rates can also indicate potential gaming or the overuse of exclusions (Machin & Sandi, 2018), and thus serve as a screening tool for further investigation. Moreover, disadvantaged and vulnerable students, such as those with special educational needs, as well as minority ethnic groups, are consistently overrepresented in school exclusion figures, a finding true across school systems around the world, including the UK (Parsons, 2018; Department for Education, 2019b), USA (Anyon, et al., 2014; Sullivan, et al., 2013) and Australia (Gardiner, et al., 1995; Hemphill, et al., 2014). The disproportionate exclusion of certain student groups emphasises the need to monitor school effects on exclusions, particularly considering the implications of institutional racism and the link between school and social exclusion (Demie, 2019). Furthermore, the entanglement of student characteristics with higher exclusion rates emphasises the importance of adjustment in order to develop 'fairer' measures of any school's unique influence.

In this article we explore how the traditional approach to school value-added models for student attainment can be extended to simultaneously incorporate multiple non-academic student outcomes and in doing so can potentially provide a more rounded approach to using student data to inform school accountability. We focus on joint modelling of student attainment, absence and exclusion data for secondary schools in England. The rest of the article is structured as follows: Section 2 describes the data; Section 3 presents the multilevel school value-added model for analysing the multiple student outcomes; Section 4 presents the results; Section 5 discusses the wider relevance of investigating non-academic student outcomes alongside academic progress for data-based school accountability systems.

## 2. Data

The data for this analysis is drawn from the National Pupil Database (NPD), an administrative dataset of all state-educated school children in England. We consider students who completed their end of secondary school General Certificate of Secondary Education (GCSE) examinations in the 2017/2018 academic year (aged 15/16), and who were attending mainstream schools. As this article is an exploration of how multiple outcomes may be examined in practice, we focus for simplicity on a random sample of 300 schools (approximately 10% of schools nationally). Schools range in size from 19 to 404 students (with a mean of 150). Across all schools, the sample consists of 45,103 students.

The three student outcomes are student attainment, absences and exclusions. Full descriptive statistics are provided in Supplementary Table S1. The student attainment outcome is derived from Attainment 8 (Department for Education, 2019c), a summary score of student performance across eight academic subjects.

The student absences outcome is defined as the total number of sessions (half days of school) a student recorded as absent during the five years of secondary schooling (the value-added period for student attainment): year 7 (age 11/12) to year 11 (age 15/16). Absences may be authorised (the majority (75%); e.g. due to illness) or unauthorised (e.g. due to truancy) though we do not pursue this distinction further here. No further detail on the reasons for each absence is presented in the data. The average student missed 89 sessions, equating to approximately 5% of their secondary schooling. Student absences were positively skewed (Supplementary Figure S1) and so in our models we log transform total absences and model the transformed response as a continuous, normally distributed outcome. An alternative approach would be to model student absences as a count or binomial count outcome (Coxe, et al., 2009; Leckie, et al., submitted) but we do not pursue this here.

The student exclusion outcome is a binary measure of whether a student has ever been excluded during secondary school: around 13% of students had been excluded one or more times. An exclusion is a formal procedure that is instigated by a school in response to behavioural incidents. This measure covers both fixed period exclusions (the overwhelming majority (99.5%)) – typically for around 2 days (Department for Education, 2019b) – and permanent exclusions, where the student must leave the establishment. As with absences, we do not have detail on the reasons for exclusion.

We adjust all three student outcomes for student prior attainment, prior absences and prior exclusion in order to account for systematic differences in these outcomes already present at school intake. Prior attainment is measured as students' combined English and maths scores at the end of primary schooling (Year 6, age 10/11). Prior attainment is standardised into a z-score (mean 0 and SD 1). Prior absence is the overall number of sessions missed by students in their final year of primary schooling. The mean number of prior absences is 15 sessions. Prior exclusion is measured by a dichotomous marker of having been excluded or not in the final year of primary schooling. Just 0.62% of students had a prior exclusion and so this measure will likely prove a less effective adjustment for school intake differences than prior attainment and prior absences.

We additionally adjust for various student background characteristics predictive of student outcomes and which vary between schools at intake (Leckie & Goldstein, 2019; Muñoz-Chereau & Thomas, 2016; Timmermans & Thomas, 2015). These include, student gender, whether a student is summer born (April to August), ethnic group (White, Black, Asian, Mixed and Other), speaking English as an additional language (EAL), Special Educational Needs (SEN) and disadvantaged status captured by eligibility for Free School Meals (FSM) in any of the previous 6 years.

Models containing school characteristics are also analysed, these include school type, school admissions policy, school gender, and whether the school has a religious denomination. Information on the categorisation of these variables is available in the Supplementary Material. Investigation of school characteristics progresses the analysis from the examination of school effects to assessment of the degree to which these may be explained by features of schools.

## 3. Methods

We estimate school value-added effects on the three student outcomes – attainment, absences and exclusions – using a multivariate, mixed-response, multilevel model. The model has a two-level structure, students (level-1) nested within schools (level-2). By jointly modelling the three outcomes we directly estimate the residual correlations between the outcome specific school effects – that is the consistency of school effects across these different dimensions of school performance. We avoid the traditional approach of fitting separate multilevel models to each outcome and correlating the predicted school effects, as this has been shown to produce biased results (Leckie, 2018).

The multivariate response model jointly estimates three equations, one for each response. Let $y_{1ij}$, $y_{2ij}$, $y_{3ij}$ denote the tri-variate response of student attainment, log absences and exclusion outcomes for student $i$ ($i = 1, \ldots, n_j$) in school $j$ ($j = 1, \ldots, J$). Student attainment and log absences are continuous outcomes which can be modelled using two-level linear regression models, whilst the exclusions outcome is binary and is modelled using a two-level probit regression model. We use probit regression in preference to the more usual logistic regression as, via its latent response formulation, it permits the correlation of the student residuals across the three equations. In this formulation of the probit model, we appeal to the notion of a unobserved continuous outcome variable $y_{3ij}^*$ which captures the propensity for the student to be excluded and underlies the observed binary outcome variable $y_{3ij}$. The relationship between $y_{3ij}$ and $y_{3ij}^*$ is simply $y_{3ij} = 1$ when $y_{3ij}^* \geq 0$ and $y_{3ij} = 0$ when $y_{3ij}^* < 0$. The model for the three continuous outcomes $y_{1ij}$, $y_{2ij}$, $y_{3ij}^*$ can then be written as follows

$$y_{1ij} = \mathbf{x}'_{1ij}\boldsymbol{\beta}_1 + u_{1j} + e_{1ij}$$

$$y_{2ij} = \mathbf{x}'_{2ij}\boldsymbol{\beta}_2 + u_{2j} + e_{2ij}$$

$$y_{3ij}^* = \mathbf{x}'_{3ij}\boldsymbol{\beta}_2 + u_{3j} + e_{3ij}$$

where $\mathbf{x}_{1ij}, \mathbf{x}_{2ij}, \mathbf{x}_{3ij}$ denote the vectors of student- and school-level covariates for the three outcomes, $\boldsymbol{\beta}_1, \boldsymbol{\beta}_2, \boldsymbol{\beta}_3$ denote the associated vectors of regression coefficients, $u_{1j}, u_{2j}, u_{3j}$ denote the school random intercept effects, and $e_{1ij}, e_{2ij}, e_{3ij}$ denote the student residuals.

The school effects and student residuals are assumed to be independent and multivariate normally distributed with zero mean vectors and unstructured covariance matrices

$$\begin{pmatrix} u_{1ij} \\ u_{2ij} \\ u_{3ij} \end{pmatrix} \sim N \left\{ \begin{pmatrix} 0 \\ 0 \\ 0 \end{pmatrix}, \begin{pmatrix} \sigma_{u_1}^2 & & \\ \sigma_{u_{12}} & \sigma_{u_2}^2 & \\ \sigma_{u_{13}} & \sigma_{u_{23}} & \sigma_{u_3}^2 \end{pmatrix} \right\}$$

$$\begin{pmatrix} e_{1ij} \\ e_{2ij} \\ e_{3ij} \end{pmatrix} \sim N \left\{ \begin{pmatrix} 0 \\ 0 \\ 0 \end{pmatrix}, \begin{pmatrix} \sigma_{e_1}^2 & & \\ \sigma_{e_{12}} & \sigma_{2_2}^2 & \\ \sigma_{e_{13}} & \sigma_{e_{23}} & 1 \end{pmatrix} \right\}$$

where the student residual variance in the equation for exclusions is constrained to a value of 1 to identify the model (the scale of the latent response $y_{3ij}^*$ is otherwise undefined).

To facilitate comparisons across the three outcomes, the results for all outcomes will be standardised onto a common response scale (with mean 0 and SD 1). The resulting regression coefficients can then be interpreted in terms of standard deviation changes in the relevant outcome variable. Results presented on the original estimation scales are provided in the Supplementary Material.

A series of four models is estimated, building in complexity to explore the utility for school accountability systems of simultaneously analysing student absences and exclusions as outcomes alongside student attainment. Model 1 is a null model without covariates, used to estimate the unadjusted school- and student-level variation and baseline correlations between outcomes. Model 2 adds prior attainment, absences and exclusions to assess a basic value-added model. Model 3 then includes the student background characteristics. Models 2 and 3 provide an increasingly adjusted version of the identified school effects, helping to isolate the independent impact of schools, with Model 3 being ultimately the most suitable for this purpose. Model 4, then adds the school characteristics, providing an appraisal of which school characteristics may relate to the outcomes and to what extent the estimated school random effects are explained by their addition.

For each model we predict, study and compare the school effects $(u_{1j}, u_{2j}, u_{3j})$ across the three outcomes. This gives a more holistic appraisal of school performance and the consistency of school effects across the three outcomes, as well as helping explore the utility of the non-academic outcomes as measures in school accountability systems.

Data preparation and descriptive statistics were carried out in Stata version 14 (StataCorp, 2015). All models were fitted using MLwiN version 3.04 (Charlton, et al., 2019) called from within Stata using the runmlwin command (Leckie & Charlton, 2013). We note that MLwiN can equally be called from within R using the sister R2MLwiN package (Zhang, et al., 2016). We fitted all models using Markov chain Monte Carlo (MCMC) methods (Browne, 2019), with diffuse prior distributions for all parameters and employing quasi-likelihood methods to provide starting values. For Models 1 to 3, the burn-in length was 500 iterations, followed by a chain of 10,000. Model 4 required a longer burn-in of 2,000 followed by a chain of 40,000 iterations. Visual assessment of the parameter chains and standard MCMC convergence diagnostics suggested that the monitored chains had converged.

## 4. Results

### *Model 1: Unadjusted model*

This first model shows the baseline variation between schools in each of the three outcomes: attainment, log absences, and exclusions (Table 1). The estimated school and student correlation matrices are presented in the Supplementary Table S2. This baseline variation captures not just differences between schools which might reasonably be attributed to variation in schools' policies and practices, but also variation due to the non-random assignment of students to schools. The Variance Partitioning Coefficient (VPC: $\sigma_{u_1}^2/\sigma_{u_1}^2 + \sigma_{e_1}^2$) (Goldstein, et al., 2002) suggests 19% (= 0.192/(0.192 + 0.808)) of total variation in student attainment lies between schools. A figure between 20 and 30% for academic outcomes is typical in studies of school effects (Grilli, et al., 2016; Reynolds, et al., 2014; Scheerens & Bosker, 1997; Teddlie & Reynolds, 2000). In contrast, the VPC for exclusions is 12% (= 0.123/(0.123 + 0.877)) and for log absences 5% (= 0.05/(0.05 + 0.95)).

These differing VPCs suggest variations in the importance of considering schools when attempting to understand each outcome. For example, most absences are authorised, for reasons such as illness or medical appointments (Department for Education, 2019a). Schools will have little control over this form of absences, and health may also be less socially graduated in adolescence (Green, 2013), serving to reduce potential differences between schools and contributing to the low VPC. Ramberg et al. (2019) found around 4% of variation occurred at the school level in a study of truancy in Stockholm, suggesting that schools may also play only a small role in unauthorised absences. The VPC for exclusions at 12% is somewhat higher. This is consistent with the notion that schools perhaps have more control over this student outcome, choosing to implement relatively more or less punitive regimes. For instance, a punitive discourse of harsh punishment dominates in the United States (Kupchik, et al., 2015) and Sullivan et al. (2013) evidenced a higher estimate of unadjusted school-level variation (29%) in student exclusions for an urban district of Wisconsin than we find here for schools in England. Given that exclusions are socially graduated, the heightened context of social segregation in the United States could have contributed to this higher variance positioned between schools, emphasising the need to adjust for student characteristics at intake using 'value-added' and 'contextual value-added' type models (Muñoz-Chereau & Thomas, 2016; Timmermans & Thomas, 2015).

Caterpillar plots for each of the three outcomes are shown in Figure 1. These plots show the predicted school effects in rank order for each response with their 95% confidence intervals. As expected, given the VPC estimates, Figure 1 illustrates that a larger proportion of schools are significantly different from average in terms of student mean attainment (69%), than for student exclusions (45%) or log absences (41%). The group of highest attaining schools in the right of the top plot are mainly grammar schools, which select their intake based on high performance in entrance examinations. We would expect the relative standing of this group to diminish in subsequent models that account for differences in prior attainment across schools.

Scatterplots and correlations of the predicted school random effects are presented in Figure 2. These show that attainment is negatively correlated with both log absences ($\rho_{u_{12}} = -0.51$) and exclusions ($\rho_{u_{13}} = -0.52$). Thus, schools whose students on average show higher attainment are likely to show lower absence and exclusion rates. In contrast, log absences and exclusions are positively correlated with one another ($\rho_{u_{23}} = 0.40$): schools where students have higher average log absences are also likely to be schools where students have a high propensity of exclusion. Student-level correlations are slightly weaker but show similar patterning and these are available in Supplementary Material. From a consistency of school effects perspective, the school correlations are nevertheless all still relatively low. Figure 2 shows considerable variability around the general associations suggested by the correlations. These results indicate that the three outcomes, whilst related, are likely to be capturing distinct aspects of schools' policies and practices, especially as some of the commonality in school outcomes is reflecting currently unmodelled school differences in student prior outcomes and other characteristics at intake. Therefore, these baseline results already emphasise the risk of a narrow viewpoint on school performance.

*Model 2: Value-added model*

The second model builds on the first through the addition of prior measures of student attainment, absences and exclusion, creating a 'value-added' model that gives a more reasonable assessment of school effects – one that adjusts for initial differences in these outcomes across schools. Table 2 shows prior attainment is strongly related to performance at the end of secondary school: a one standard deviation (SD) increase in prior attainment relates to a 0.674 SD increase in final attainment. There is also a considerable drop in the school-level variance for attainment between Models 1 and 2 (0.192 to 0.071) showing that approximately two-thirds of the variation identified between schools for attainment can be attributed to differences in prior attainment at intake. Caterpillar plots for Model 2 (Figure 3) demonstrate that there no longer appears to be an outlying group of high achieving schools in the top plot; by including prior attainment in the model we have controlled for a major selection effect exhibited by grammar schools, thus bringing their measured performance more in line with other schools. Higher prior attainment also relates to fewer log absences, and a lower propensity of exclusion. However, as expected given the correlations in Figure 2, the magnitude of these effects is much lower than it was for final attainment: a 1SD increase in prior attainment is associated with a 0.123 SD decrease in log absences and a 0.230 SD decrease in the propensity to be excluded.

Prior absences are associated strongly with increased log absences throughout secondary school: an additional week (10 sessions) of absence in the final year of primary school is associated with an expected increase of 0.30 SD in log absences. The effect of prior absences on attainment score and propensity of exclusion are relatively small in comparison.

An additional week of prior absences is associated with an expected decrease of 0.05 SD in final attainment score and an increased propensity to be excluded of 0.06 SD. Note that these effects should not be interpreted as causal as the prior measures to which they relate will in part be proxying for omitted student characteristics and other factors which are correlated with both the modelled student outcomes and their prior measures.

The inclusion of prior absences makes an important contribution to explaining school level and total variation in log absences: the school-level variance declines by over a quarter from 0.050 to 0.036, and the R-Squared is 0.24, showing approximately 24% of total variation is explained by the fixed part of Model 2. In models containing only prior attainment or only prior exclusion (results available in Supplementary Material) the R-Squared for log absences was 0.040 and 0.003 respectively. Therefore, if data on student absences is to be evaluated within systems of school accountability it would appear crucial to consider a schema of 'value-added' models, like the treatment of attainment measures in the creation of progress scores.

Having been excluded in the final year of primary school is associated to all three outcomes: predicting a large 0.546 SD reduction in attainment, a 0.277 SD increase in log absences and a 1.221 SD increase in the propensity of exclusion throughout secondary school. However, prior exclusion makes relatively little difference to the school-level variance or the R-Squared estimate. This is likely because so few students are excluded in the final year of primary school (<1%). A longer history of exclusions over the whole of primary school may be a more appropriate measure of prior circumstance but we do not have these data available. Additionally, exclusions are more generally a rarer phenomenon in primary rather than secondary schooling.

Comparing the scatterplots and correlations in Figure 4 based on the current model to those in Figure 2 for the previous model demonstrates that school intake differences in prior attainment, absences and exclusion were driving a similarity in the three later outcomes. The school-level correlations between the adjusted school effects decrease substantially, for example, the correlation between school effects on student attainment and log absences drops from -0.51 in Model 1 to -0.24 in Model 2. Information on school effects on student attainment may, therefore, not provide much information regarding school influences on non-academic outcomes such as student absences or exclusions: a singular school performance measure is inadequate to capture these broader dimensions of educational performance.

*Model 3: Contextual value-added model*

Model 3 adds additional student background characteristics, further adjusting school differences on each student outcome to achieve a fairer comparison of schools. This model also allows evaluation of how student sociodemographic factors relate to the three outcomes and the identification of any notable divergences. The results are presented in Table 3 with estimated correlation matrices in Supplementary Table S4.

The school-level variances for attainment and exclusions exhibit further declines from Model 2 to Model 3, showing that school differences in the sociodemographic characteristics of their students at intake impact on average academic attainment and average propensity of exclusion, even after adjusting for prior versions of these outcomes. For log absences, the school-level variance remains around the same level, and the correlation between the school effects for log absences between Model 2 and 3 is 0.96. This could suggest that the need to adjust for student sociodemographics in order to achieve fairer school comparisons is less potent for log absences. However, the R-Squared statistic shows that explanatory power is nonetheless provided by the addition of student background characteristics (increasing from 0.24 to 0.28).

The school effects from Model 3, arguably the most suitable for the purposes of school accountability, make it further evident that schools often have very different effects on different student outcomes. Figure 5 shows that the correlations between outcomes have again decreased, for instance that between attainment and log absences at the school-level drops from -0.24 in Model 2 to -0.12 in Model 3. These plots nonetheless retain a potential utility as screening tools, helping identify unusual outlying cases where a more in-depth follow-up may be prudent. For instance, the plots show two schools with particularly low rates of log absences even after adjustments for prior absence and student background characteristics.

The fixed-part coefficients in Table 3 demonstrate that many of the associations for log absences and exclusions are in the opposite direction to that for attainment. This is expected given the negative school- and student-level correlations seen in Model 1. Notable exceptions to this general patterning are shown by gender and for the Black and Mixed ethnic groups. Girls have a lower propensity of exclusion (-0.354 SD) but higher log absences (0.069 SD) than boys. Women often report worse health than men, so this is likely due to greater illness absences. The Black ethnic group has a positive coefficient for attainment (0.039 SD) and they are estimating to record fewer log absences (-0.415 SD), but they have a markedly higher propensity of exclusion (0.361 SD). UK literature highlights that it is Black Caribbean students who tend to suffer a greater burden of exclusions (Department for Education, 2019b; Parsons, 2018), whilst Black African students tend to perform better academically (Leckie & Goldstein, 2019). Such underlying trends are likely playing out beneath the overall 'Black' category and balancing out the results found here.

*Model 4: Contextual value-added model with school characteristics*

The final model, presented in Table 4, examines how the predicted school effects systematically vary across a range of key structural school characteristics. The school-level variance reveals only a marginal decline for log absences, and the lack of explanatory power is further emphasised by the R-Squared figure which does not improve between Models 3 and 4 for log absences. In contrast, the school-level variance for attainment decreases from 0.059 to 0.037, and for exclusions from 0.086 to 0.076, demonstrating systematic differences in school effects on student outcomes across the considered school characteristics.

From Table 4 it appears that students in more technically or vocationally orientated schools (Studio schools or University Technical Colleges (UTCs)) are expected to score 0.306 SD lower in attainment than more general community schools, after taking account of student characteristics. This highlights another potential flaw in systems of school accountability that take a 'one-size-fits-all' approach; measures based solely on traditional academic performance will disadvantage those educational establishments aiming to provide a broader set of skills and qualifications. The inclusion of non-academic outcomes may help compensate against this.

The results also demonstrate that those in grammar schools are predicted to score 0.433 SD higher than those in non-selective schools but are also expected to show higher log absences (0.156 SD). It may be that the type of absences students take in grammar schools are not those that negatively affect performance, for instance, study leave or holidays which could confer other formative benefits to learning. Therefore, for certain types of schools, a high absence rate may not necessarily exist alongside negative academic performance and care is required in the interpretation of the results.

## 5. Discussion

School accountability systems around the world increasingly rely on school value-added models derived from student attainment data but limit their scope to a few or singular academic measures, giving a narrow viewpoint on the contribution of schools to student outcomes. In this article we explored how the traditional multilevel approach to value-added models can be extended to simultaneously analyse a range of academic and non-academic outcomes and to, therefore, reflect a more holistic perspective on the role of schools. We focused on student attainment, absences and exclusions in secondary schools in England.

Our main finding is that these three outcomes capture different aspects of school effectiveness. This was particularly evident in the 'contextual value-added' model, often argued to be the most appropriate for making fair and meaningful comparisons between schools (Leckie & Goldstein, 2017). In this model, the correlation between the school effects for student attainment and log absences, for example, was just -0.12. This finding aligns with that of Smyth (1999, p. 488) who estimated using Irish data a correlation of -0.28 in their study of school effects on student exam performance and chronic absenteeism. More generally, the results for the specific schools in our data show many instances where school results are patterned contrary to the expected relationship. For example, instances with higher than expected attainment that nonetheless show higher than expected absence and exclusion rates. Schools which are effective on one dimension of school performance are not guaranteed to be effective on other dimensions. A singular focus on attainment is clearly not sufficient to capture the true breadth of schools' contribution to student learning, performance and development. A more holistic appraisal of schools for the purpose of accountability would be gained from considering a broader range of student outcomes.

There are long-standing concerns with the use of high-stakes tests in school accountability systems, especially deleterious consequences such as curriculum narrowing, cultures of fear and gaming behaviour (Amrein-Beardsley, 2014; Foley & Goldstein, 2012). Broadening the scope of student outcomes analysed in school value-added models and accountability systems, and thus lessening the stakes attached to singular or very few tests, may potentially alleviate some of these negative side-effects. For example, if exclusions were considered a key outcome for schools, this could deter the strategic use of exclusions to re-shape the test pool (Ehren & Swanborn, 2012). Moreover, a schema for school accountability which additionally considers non-academic student outcomes would likely prove helpful when identifying schools for closer inspection in the name of school improvement. For instance, a set percentage of schools from the tails of the caterpillar plots could be picked for inspection to better understand extreme outcomes. Similarly, scatterplots of school effects on different outcomes could form part of an informed screening tool to reveal schools which only prove unusual in their combination of effects. The introduction of school characteristics can then potentially flag groups of schools which appear to collectively be operating in atypical ways. For example, in the present analyses, grammar were schools were shown to be effective in raising student attainment but ineffective in controlling student absence rates and these results persisted even after adjustment for a rich range of student background characteristics.

There are various limitations to our work, and we encourage researchers to pursue these in further research. First, for log absences and exclusions the total variance explained was markedly lower than that achieved for attainment (56% for attainment versus 28% and 17% for log absences and exclusions respectively in the final 'contextual value-added' model). For attainment most of the explanatory power was carried by the measure of prior attainment. For absences and exclusions, we were restricted to using prior data from the last year of primary school as opposed to data spanning the entirety of students' primary schooling and this may explain the differential explanatory power of the prior measures across the three outcomes. Additionally, in contrast to student attainment, student absence and exclusion data are recorded continuously during students' schooling. While we have incorporated this new form of student outcome data in a parallel way to student attainment, this is just one possible approach. We encourage others to continue to consider how best to enact value-added models with continuously recorded student outcome data.

Another limitation is that for this analysis we focused on overall summaries of absences and exclusions, where authorised and unauthorised absences, and permanent and fixed period exclusions, were combined. There may well be substantive differences in the effect of schools on these different facets of absences and exclusions. For example, schools may have more influence on unauthorised absences, or truancy, which may be due to a breakdown of student-school relationships, or negative school experiences (Gottfried, 2019; Ramberg, et al., 2019). Distinguishing between authorised and unauthorised absences, or permanent and fixed period exclusions, would also help isolate which elements of these non-academic outcomes have the strongest correlations with attainment. Further research would be needed to uncover these extra dimensions of school effectiveness and the implications of these categorisations for the potential use of such outcomes in systems of school accountability. One related complication to permanent exclusions is that these imply a subsequent change of schools; we have not explored any such student mobility in the current analyses. School value-added models can be extended to account for student mobility (Goldstein, et al., 2007; Leckie, 2009) and we encourage further work in this area with respect to modelling student exclusions in particular.

Perhaps the most important limitation of our work is that student absences and exclusions are just two of many potential non-academic outcomes which one might explore in a more holistic approach to monitoring school performance. Other measures such as student destinations may also provide insight into further aspects of school effectiveness. Our model naturally extends to the inclusion of a fourth and further student outcomes where these are available. However, possibly most significantly, there are many important student outcomes such as student engagement and aspirations which rarely appear in national administrative datasets and are far from straightforward to measure. Further work is needed to explore new ways to incorporate such outcomes if we are to start to address the weaknesses of current systems of school accountability.

Table 1. Model 1 (unadjusted model) regression coefficients and variance components.

|  | Attainment | | Log Absences | | Exclusions | |
| --- | --- | --- | --- | --- | --- | --- |
|  | Est | SE | Est | SE | Est | SE |
| *Fixed* | | | | | | |
| Intercept | 2.449 | (0.026) | 4.269 | (0.014) | -1.099 | (0.022) |
| *Random* | | | | | | |
| School variance | 0.192 | (0.016) | 0.050 | (0.005) | 0.123 | (0.012) |
| Student variance | 0.808 | (0.005) | 0.950 | (0.006) | 0.877 | . |
| *Statistics* | | | | | | |
| VPC | 19% | | 5% | | 12% | |
| R-Squared | 0.00 | | 0.00 | | 0.00 | |

Note. The results are standardised onto a common response scale (with mean 0 and SD 1) to facilitate comparisons across the three student outcomes. Full estimated correlation matrices in Supplementary Table S2.

Table 2. Model 2 (value-added model) regression coefficients and variance components.

|  | Attainment | | Log Absences | | Exclusions | |
| --- | --- | --- | --- | --- | --- | --- |
|  | Est | SE | Est | SE | Est | SE |
| *Fixed* | | | | | | |
| Intercept | 2.649 | (0.016) | 3.806 | (0.012) | -1.203 | (0.022) |
| Prior attainment | 0.674 | (0.003) | -0.123 | (0.004) | -0.230 | (0.007) |
| Prior absences | -0.005 | (<0.001) | 0.030 | (<0.001) | 0.006 | (<0.001) |
| Prior exclusion | -0.546 | (0.041) | 0.277 | (0.050) | 1.221 | (0.073) |
| *Random* | | | | | | |
| School variance | 0.071 | (0.006) | 0.036 | (0.004) | 0.102 | (0.011) |
| Student variance | 0.446 | (0.003) | 0.723 | (0.005) | 0.819 | . |
| *Statistics* | | | | | | |
| VPC | 14% | | 5% | | 11% | |
| R-Squared | 0.48 | | 0.24 | | 0.08 | |

Note. The results are standardised onto a common response scale (with mean 0 and SD 1) to facilitate comparisons across the three student outcomes. Full estimated correlation matrices in Supplementary Table S3.

Table 3. Model 3 (contextual value-added model) regression coefficients and variance components.

|  | Attainment | | Log Absences | | Exclusions | |
|---|---|---|---|---|---|---|
|  | Est | SE | Est | SE | Est | SE |
| *Fixed* | | | | | | |
| Intercept | 2.557 | (0.016) | 3.733 | (0.014) | -1.111 | (0.023) |
| Prior attainment | 0.641 | (0.004) | -0.093 | (0.005) | -0.186 | (0.008) |
| Prior absences | -0.004 | (<0.001) | 0.028 | (<0.001) | 0.004 | (<0.001) |
| Prior exclusion | -0.382 | (0.039) | 0.242 | (0.050) | 0.973 | (0.076) |
| Summer born | 0.042 | (0.006) | -0.037 | (0.008) | -0.072 | (0.014) |
| Female | 0.208 | (0.006) | 0.069 | (0.008) | -0.354 | (0.015) |
| Black | 0.039 | (0.016) | -0.415 | (0.020) | 0.361 | (0.032) |
| Asian | 0.123 | (0.014) | -0.065 | (0.017) | -0.033 | (0.032) |
| Mixed | 0.054 | (0.015) | -0.038 | (0.020) | 0.207 | (0.033) |
| Other | 0.105 | (0.023) | -0.052 | (0.029) | -0.006 | (0.049) |
| EAL | 0.251 | (0.011) | -0.073 | (0.015) | -0.135 | (0.026) |
| SEN | -0.182 | (0.010) | 0.105 | (0.013) | 0.159 | (0.020) |
| FSM | -0.272 | (0.007) | 0.368 | (0.010) | 0.348 | (0.016) |
| *Random part parameters* | | | | | | |
| School variance | 0.059 | (0.005) | 0.037 | (0.004) | 0.086 | (0.009) |
| Student variance | 0.414 | (0.003) | 0.685 | (0.004) | 0.765 | . |
| *Statistics* | | | | | | |
| VPC | 12% | | 5% | | 10% | |
| R-Squared | 0.53 | | 0.28 | | 0.15 | |

Note. The results are standardised onto a common response scale (with mean 0 and SD 1) to facilitate comparisons across the three student outcomes. Full estimated correlation matrices in Supplementary Table S4.

Table 4. Model 4 (contextual value-added model with school characteristics) regression coefficients and variance components.

|  | Attainment | | Log Absences | | Exclusions | |
| --- | --- | --- | --- | --- | --- | --- |
|  | Est | SE | Est | SE | Est | SE |
| *Fixed* | | | | | | |
| Intercept | 2.416 | (0.025) | 3.756 | (0.024) | -1.076 | (0.039) |
| Prior attainment | 0.628 | (0.003) | -0.094 | (0.005) | -0.184 | (0.008) |
| Prior absences | -0.004 | (<0.001) | 0.028 | (<0.001) | 0.004 | (<0.001) |
| Prior exclusion | -0.374 | (0.038) | 0.244 | (0.051) | 0.968 | (0.071) |
| Summer born | 0.041 | (0.006) | -0.036 | (0.008) | -0.072 | (0.014) |
| Female | 0.205 | (0.006) | 0.070 | (0.008) | -0.348 | (0.015) |
| Black | 0.037 | (0.015) | -0.416 | (0.020) | 0.361 | (0.031) |
| Asian | 0.118 | (0.014) | -0.067 | (0.018) | -0.024 | (0.032) |
| Mixed | 0.051 | (0.015) | -0.038 | (0.019) | 0.210 | (0.031) |
| Other | 0.101 | (0.022) | -0.052 | (0.029) | 0.000 | (0.051) |
| EAL | 0.245 | (0.011) | -0.072 | (0.015) | -0.133 | (0.026) |
| SEN | -0.179 | (0.010) | 0.104 | (0.013) | 0.157 | (0.020) |
| FSM | -0.267 | (0.007) | 0.369 | (0.010) | 0.343 | (0.016) |
| Academy type | 0.116 | (0.028) | -0.034 | (0.027) | -0.105 | (0.042) |
| Sponsored academy | -0.008 | (0.034) | -0.059 | (0.032) | 0.108 | (0.052) |
| Studio/UTC | -0.306 | (0.105) | 0.381 | (0.115) | 0.236 | (0.173) |
| Grammar | 0.433 | (0.064) | 0.156 | (0.068) | -0.006 | (0.111) |
| Secondary modern | -0.007 | (0.059) | 0.100 | (0.061) | 0.090 | (0.090) |
| Boys | 0.082 | (0.060) | 0.025 | (0.058) | -0.044 | (0.094) |
| Girls | 0.122 | (0.053) | 0.028 | (0.051) | -0.160 | (0.086) |
| Religious | 0.081 | (0.030) | -0.026 | (0.031) | -0.019 | (0.047) |
| *Random part parameters* | | | | | | |
| School variance | 0.037 | (0.003) | 0.035 | (0.003) | 0.076 | (0.008) |
| Student variance | 0.399 | (0.003) | 0.688 | (0.005) | 0.754 | . |
| *Statistics* | | | | | | |
| VPC | 9% | | 5% | | 9% | |
| R-Squared | 0.56 | | 0.28 | | 0.17 | |

Note. The results are standardised onto a common response scale (with mean 0 and SD 1) to facilitate comparisons across the three student outcomes. Full estimated correlation matrices in Supplementary Table S5.

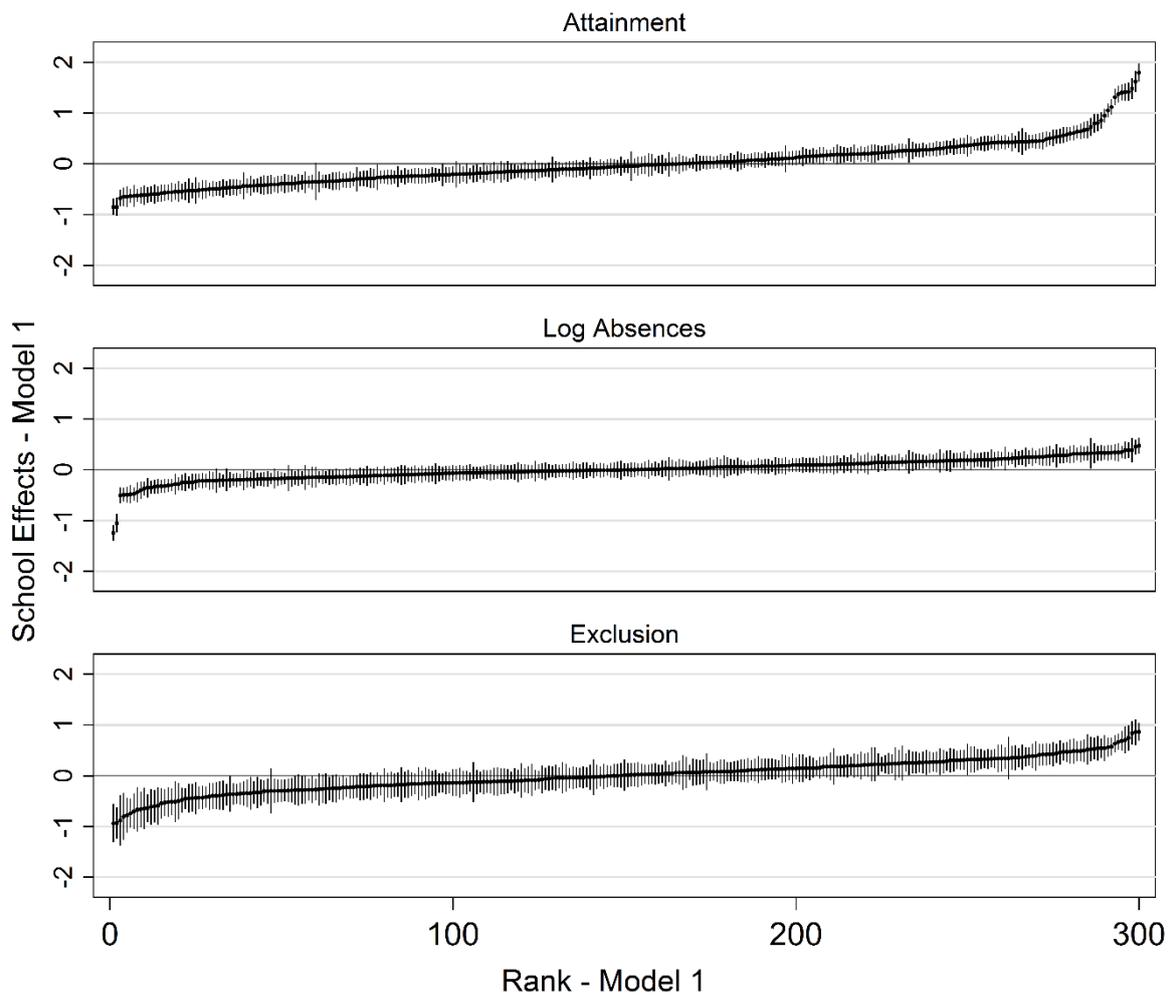

Figure 1. Caterpillar plots of predicted school effects for Model 1: unadjusted model.

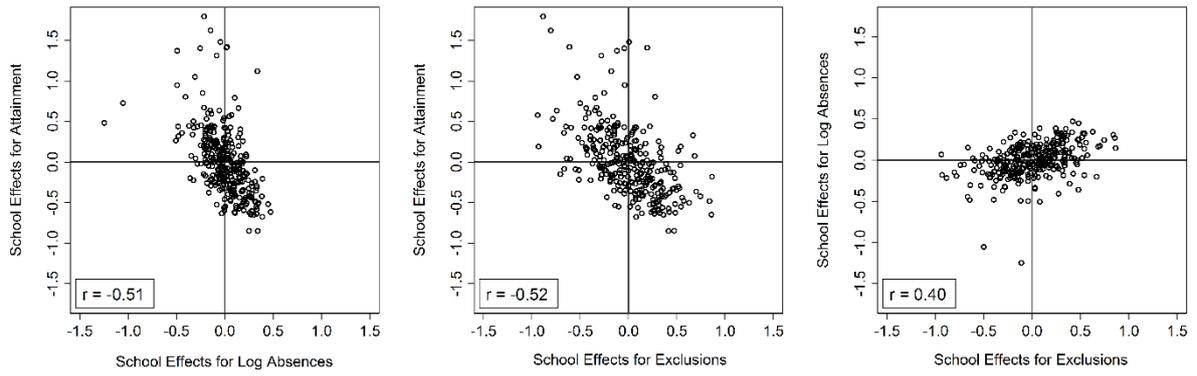

Figure 2. Scatterplots of predicted school effects between attainment, log absences and exclusions for Model 1: unadjusted model.

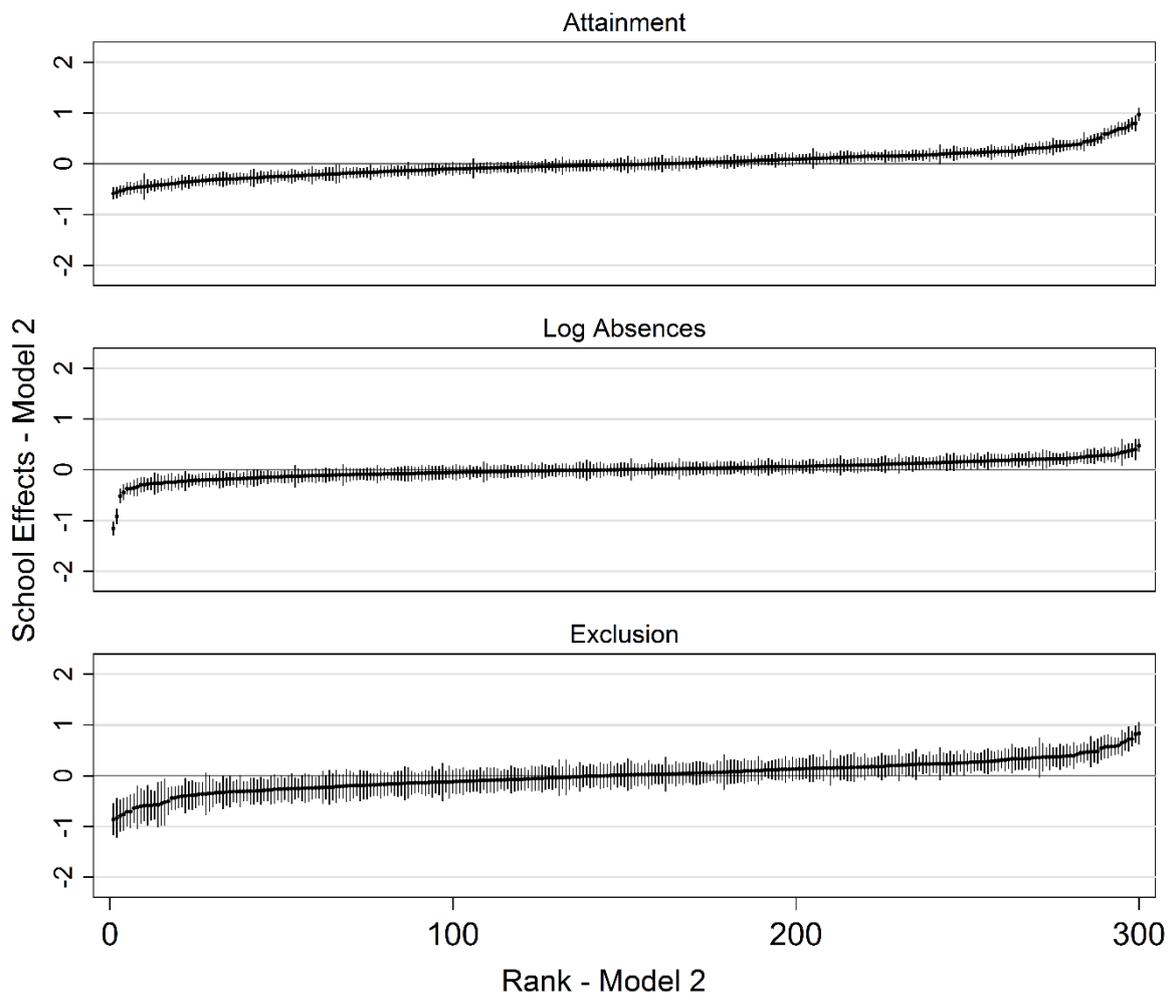

Figure 3. Caterpillar plots of predicted school effects for Model 2: value-added model.

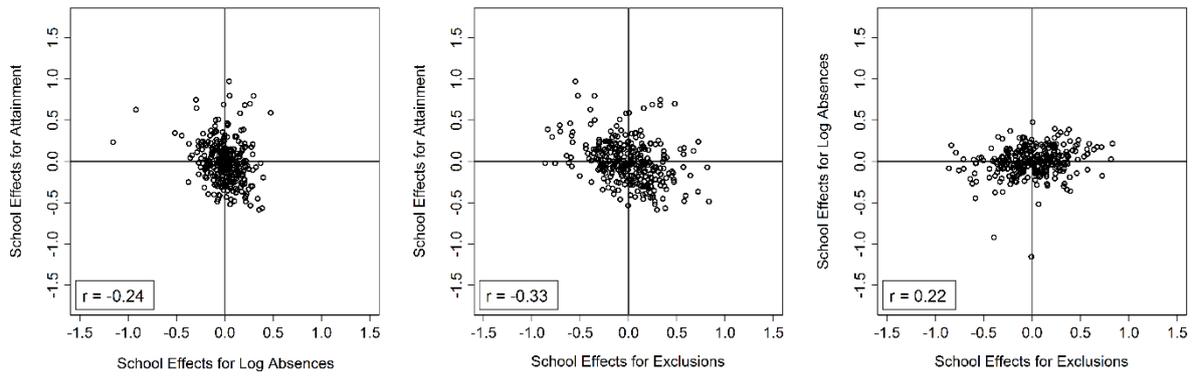

Figure 4. Scatterplots of predicted school effects between attainment, log absences and exclusions for Model 2: value-added model.

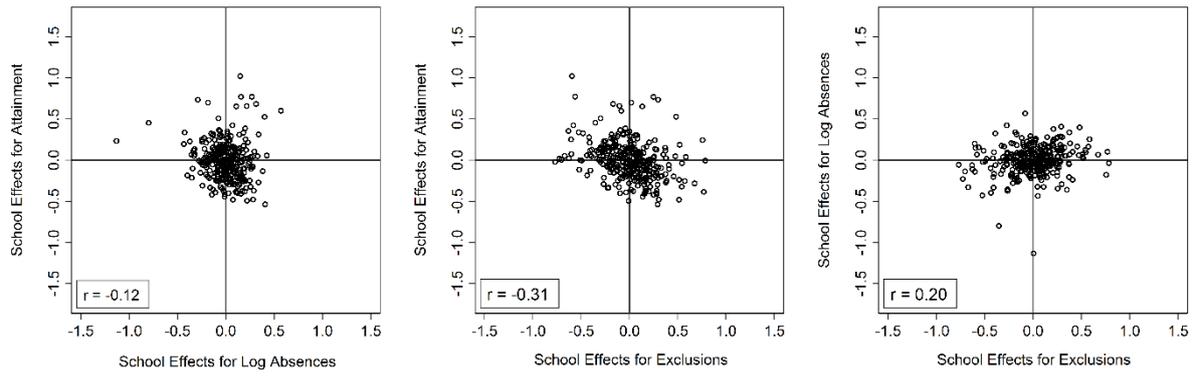

Figure 5. Scatterplots of predicted school effects between attainment, log absences and exclusions for Model 3: contextual value-added model.

**Supplementary Material**

Supplementary Table S1. Descriptive statistics.

|  |  | Mean | SD | % of Students | Number of Students |
|---|---|---|---|---|---|
| *Responses* | | | | | |
| Attainment 8 | | 47.36 | 18.88 | . | 45,103 |
| Total Absences | | 88.61 | 92.56 | . | 45,103 |
| Log Absences | | 4.10 | 0.96 | . | 45,103 |
| Exclusion | Not Excluded | . | . | 87.09 | 39,278 |
| | Excluded | . | . | 12.91 | 5,825 |
| *Pupil Characteristics* | | | | | |
| KS2 Score | | 4.76 | 0.71 | . | 45,103 |
| Prior Attainment | | 0.00 | 1.00 | . | 45,103 |
| Prior Absences | | 15.46 | 15.22 | . | 45,103 |
| Prior Exclusion | Not Excluded | . | . | 99.38 | 44,825 |
| | Excluded | . | . | 0.62 | 278 |
| Summer Born | Not Summer Born | . | . | 57.75 | 26,046 |
| | Summer Born | . | . | 42.25 | 19,057 |
| Gender | Male | . | . | 51.43 | 23,195 |
| | Female | . | . | 48.57 | 21,908 |
| Ethnic Group | White | . | . | 76.92 | 34,692 |
| | Black | . | . | 5.66 | 2,554 |
| | Asian | . | . | 10.75 | 4,847 |
| | Mixed | . | . | 4.62 | 2,085 |
| | Other | . | . | 2.05 | 925 |
| English as Additional Language (EAL) | Not EAL | . | . | 85.45 | 38,539 |
| | EAL | . | . | 14.55 | 6,564 |
| Special Educational Needs (SEN) | No SEN | . | . | 87.21 | 39,333 |
| | SEN | . | . | 12.79 | 5,770 |
| Free Schools Meals (FSM) | Not eligible for FSM in last 6 years | . | . | 73.74 | 33,261 |
| | Eligible for FSM in last 6 years | . | . | 26.26 | 11,842 |

Supplementary Table S1 continued. Descriptive statistics.

| School Characteristics | | % of Schools | Number of Schools | % of Students | Number of Students |
|---|---|---|---|---|---|
| School Type | Community type | 31.00 | 93 | 31.11 | 14,030 |
| | Academy type | 43.67 | 131 | 46.70 | 21,061 |
| | Sponsored Academy | 24.00 | 72 | 21.69 | 9,784 |
| | Studio/University Technical College | 1.33 | 4 | 0.51 | 228 |
| Admissions Policy | Comprehensive | 91.00 | 273 | 92.45 | 41,699 |
| | Grammar | 5.00 | 15 | 3.88 | 1,751 |
| | Secondary Modern | 4.00 | 12 | 3.66 | 1,653 |
| School Gender | Mixed | 88.00 | 264 | 89.90 | 40,546 |
| | Boys | 6.67 | 20 | 4.86 | 2,190 |
| | Girls | 5.33 | 16 | 5.25 | 2,367 |
| Religious Character | No religious character | 81.00 | 243 | 83.53 | 37,676 |
| | Religious | 19.00 | 57 | 16.47 | 7,427 |

Note on school characteristics:

- School type
  - *Community type (reference category)*: These are school types which are under Local Authority control which must follow the national curriculum. School types included in this category are Community, Foundation, Voluntary Aided, and Voluntary Controlled.
  - *Academy type*: These are schools which are funded directly by the government with more freedom over management and curriculum. School types included in this category are Converter academies and Free schools.
  - *Sponsored academy*: This is a subset of academy schools primarily composed on schools which were disadvantaged pre-conversion.
  - *Studio/UTC*: Studio schools and University Technical Colleges (UTCs) are more specialist, with a technical/vocational focus and provide education from 14 years of age rather than the beginning of secondary education.
  - See https://www.gov.uk/types-of-school for more information on school types in England.
- School admissions policy
  - *Comprehensive (reference category)*: non-selective.
  - *Grammar*: selective.
  - *Secondary modern*: non-selective in an area serving selective schools.

Supplementary Table S2. Model 1 (unadjusted model) school- and student-level correlations between outcomes.

| *Level 2: School* | Attainment | Log Absences | Exclusions |
|---|---|---|---|
| Attainment | 1 | | |
| Log Absences | -0.513 | 1 | |
| Exclusions | -0.515 | 0.401 | 1 |
| *Level 1: Pupil* | Attainment | Log Absences | Exclusions |
| Attainment | 1 | | |
| Log Absences | -0.390 | 1 | |
| Exclusions | -0.501 | 0.397 | 1 |

Supplementary Table S3. Model 2 (value-added model) school- and student-level correlations between outcomes.

| Level 2: School | Attainment | Log Absences | Exclusions |
|---|---|---|---|
| Attainment | 1 | | |
| Log Absences | -0.235 | 1 | |
| Exclusions | -0.329 | 0.216 | 1 |
| Level 1: Pupil | Attainment | Log Absences | Exclusions |
| Attainment | 1 | | |
| Log Absences | -0.353 | 1 | |
| Exclusions | -0.465 | 0.363 | 1 |

Supplementary Table S4. Model 3 (contextual value-added model) school- and student-level correlations between outcomes.

| *Level 2: School* | Attainment | Log Absences | Exclusions |
|---|---|---|---|
| Attainment | 1 | | |
| Log Absences | -0.116 | 1 | |
| Exclusions | -0.306 | 0.200 | 1 |
| *Level 1: Pupil* | Attainment | Log Absences | Exclusions |
| Attainment | 1 | | |
| Log Absences | -0.342 | 1 | |
| Exclusions | -0.429 | 0.370 | 1 |

Supplementary Table S5. Model 4 (contextual value-added model) school- and student-level correlations between outcomes.

| *Level 2: School* | Attainment | Log Absences | Exclusions |
|---|---|---|---|
| Attainment | 1 | | |
| Log Absences | -0.224 | 1 | |
| Exclusions | -0.220 | 0.237 | 1 |
| *Level 1: Pupil* | Attainment | Log Absences | Exclusions |
| Attainment | 1 | | |
| Log Absences | -0.342 | 1 | |
| Exclusions | -0.429 | 0.370 | 1 |

Supplementary Table S6. Model 1 (unadjusted model) unstandardized regression coefficients and variance components.

|  | Attainment | | Log Absences | | Exclusions | |
| --- | --- | --- | --- | --- | --- | --- |
|  | Est | SE | Est | SE | Est | SE |
| *Fixed* | | | | | | |
| Intercept | 47.018 | (0.494) | 4.100 | (0.013) | -1.173 | (0.023) |
| *Random* | | | | | | |
| School Variance | 70.761 | (6.042) | 0.046 | (0.004) | 0.140 | (0.014) |
| Student Variance | 297.752 | (2.018) | 0.877 | (0.006) | 1.000 | . |

Supplementary Table S7. Model 2 (value-added model) unstandardized regression coefficients and variance components.

|  | Attainment | | Log Absences | | Exclusions | |
|---|---|---|---|---|---|---|
|  | Est | SE | Est | SE | Est | SE |
| *Fixed* | | | | | | |
| Intercept | 48.894 | (0.295) | 3.649 | (0.012) | -1.330 | (0.024) |
| Prior attainment | 12.431 | (0.061) | -0.118 | (0.004) | -0.254 | (0.008) |
| Prior absences | -0.099 | (0.004) | 0.029 | (<0.001) | 0.006 | (<0.001) |
| Prior exclusion | -10.079 | (0.757) | 0.265 | (0.048) | 1.349 | (0.081) |
| *Random* | | | | | | |
| School Variance | 24.129 | (2.058) | 0.033 | (0.003) | 0.125 | (0.013) |
| Student Variance | 151.978 | (1.039) | 0.664 | (0.004) | 1.000 | . |

Supplementary Table S8. Model 3 (contextual value-added model) unstandardized regression coefficients and variance components.

|  | Attainment | | Log Absences | | Exclusions | |
|---|---|---|---|---|---|---|
|  | Est | SE | Est | SE | Est | SE |
| *Fixed* | | | | | | |
| Intercept | 47.051 | (0.286) | 3.597 | (0.014) | -1.271 | (0.027) |
| Prior attainment | 11.798 | (0.066) | -0.090 | (0.004) | -0.213 | (0.009) |
| Prior absences | -0.072 | (0.004) | 0.027 | (<0.001) | 0.005 | (0.001) |
| Prior exclusion | -7.036 | (0.722) | 0.233 | (0.048) | 1.112 | (0.087) |
| Summer born | 0.767 | (0.113) | -0.035 | (0.008) | -0.082 | (0.016) |
| Female | 3.828 | (0.118) | 0.066 | (0.008) | -0.405 | (0.017) |
| Black | 0.725 | (0.287) | -0.399 | (0.019) | 0.413 | (0.037) |
| Asian | 2.258 | (0.261) | -0.063 | (0.017) | -0.037 | (0.037) |
| Mixed | 0.990 | (0.282) | -0.037 | (0.019) | 0.237 | (0.037) |
| Other | 1.923 | (0.419) | -0.050 | (0.028) | -0.007 | (0.056) |
| EAL | 4.611 | (0.209) | -0.070 | (0.014) | -0.155 | (0.030) |
| SEN | -3.347 | (0.185) | 0.101 | (0.012) | 0.182 | (0.023) |
| FSM | -5.013 | (0.136) | 0.354 | (0.009) | 0.398 | (0.018) |
| *Random* | | | | | | |
| School Variance | 20.009 | (1.745) | 0.035 | (0.003) | 0.113 | (0.012 |
| Student Variance | 140.262 | (0.916) | 0.636 | (0.004) | 1.000 | . |

Supplementary Table S9. Model 4 (contextual value-added model with school characteristics) unstandardized regression coefficients and variance components.

|  | Attainment | | Log Absences | | Exclusions | |
|---|---|---|---|---|---|---|
|  | Est | SE | Est | SE | Est | SE |
| *Fixed* | | | | | | |
| Intercept | 45.293 | (0.462) | 3.611 | (0.023) | -1.238 | (0.045) |
| Prior attainment | 11.777 | (0.065) | -0.091 | (0.004) | -0.212 | (0.009) |
| Prior absences | -0.072 | (0.004) | 0.027 | (<0.001) | 0.005 | (0.001) |
| Prior exclusion | -7.014 | (0.717) | 0.234 | (0.049) | 1.115 | (0.082) |
| Summer born | 0.764 | (0.117) | -0.035 | (0.008) | -0.083 | (0.016) |
| Female | 3.836 | (0.120) | 0.068 | (0.008) | -0.401 | (0.017) |
| Black | 0.700 | (0.285) | -0.400 | (0.019) | 0.416 | (0.036) |
| Asian | 2.220 | (0.257) | -0.064 | (0.017) | -0.028 | (0.036) |
| Mixed | 0.959 | (0.278) | -0.036 | (0.019) | 0.242 | (0.036) |
| Other | 1.888 | (0.420) | -0.050 | (0.028) | -0.001 | (0.058) |
| EAL | 4.595 | (0.209) | -0.069 | (0.014) | -0.153 | (0.029) |
| SEN | -3.357 | (0.186) | 0.100 | (0.013) | 0.181 | (0.023) |
| FSM | -5.001 | (0.139) | 0.355 | (0.009) | 0.396 | (0.018) |
| Academy type | 2.178 | (0.519) | -0.033 | (0.026) | -0.121 | (0.049) |
| Sponsored academy | -0.151 | (0.636) | -0.057 | (0.031) | 0.124 | (0.060) |
| Studio/UTC | -5.731 | (1.977) | 0.366 | (0.111) | 0.272 | (0.200) |
| Grammar | 8.112 | (1.207) | 0.150 | (0.066) | -0.007 | (0.128) |
| Secondary modern | -0.140 | (1.111) | 0.096 | (0.058) | 0.104 | (0.104) |
| Boys | 1.535 | (1.128) | 0.024 | (0.056) | -0.051 | (0.108) |
| Girls | 2.286 | (0.992) | 0.027 | (0.049) | -0.184 | (0.100) |
| Religious denomination | 1.515 | (0.566) | -0.025 | (0.030) | -0.022 | (0.054) |
| *Random* | | | | | | |
| School Variance | 13.104 | (1.177) | 0.032 | (0.003) | 0.101 | (0.011) |
| Student Variance | 140.268 | (0.942) | 0.636 | (0.004) | 1.000 | . |

Supplementary Table S10. Value-added model with prior attainment only regression coefficients and variance components.

|  | Attainment | | Log Absences | | Exclusions | |
| --- | --- | --- | --- | --- | --- | --- |
|  | Est | SE | Est | SE | Est | SE |
| *Fixed* | | | | | | |
| Intercept | 2.560 | (0.016) | 4.271 | (0.013) | -1.111 | (0.021) |
| Prior attainment | 0.687 | (0.003) | -0.192 | (0.005) | -0.247 | (0.007) |
| *Random* | | | | | | |
| School Variance | 0.075 | (0.006) | 0.043 | (0.004) | 0.104 | (0.011) |
| Student Variance | 0.454 | (0.003) | 0.920 | (0.006) | 0.834 | . |
| *Statistics* | | | | | | |
| VPC | 14% | | 4% | | 11% | |
| R-Squared | 0.47 | | 0.04 | | 0.06 | |

Note. The results are standardised onto a common response scale (with mean 0 and SD 1) to facilitate comparisons across the three student outcomes.

Supplementary Table S11. Value added model with prior absences only regression coefficients and variance components.

|  | Attainment | | Log Absences | | Exclusions | |
|---|---|---|---|---|---|---|
|  | Est | SE | Est | SE | Est | SE |
| *Fixed* | | | | | | |
| Intercept | 2.644 | (0.025) | 3.796 | (0.013) | -1.228 | (0.022) |
| Prior absences | -0.011 | (<0.001) | 0.031 | (<0.001) | 0.008 | (<0.001) |
| *Random* | | | | | | |
| School Variance | 0.179 | (0.015) | 0.038 | (0.004) | 0.116 | (0.012) |
| Student Variance | 0.792 | (0.005) | 0.739 | (0.005) | 0.869 | . |
| *Statistics* | | | | | | |
| VPC | 18% | | 5% | | 12% | |
| R-Squared | 0.03 | | 0.22 | | 0.02 | |

Note. The results are standardised onto a common response scale (with mean 0 and SD 1) to facilitate comparisons across the three student outcomes.

Supplementary Table S12. Value added model with prior exclusion only regression coefficients and variance components.

|  | Attainment | | Log Absences | | Exclusions | |
| --- | --- | --- | --- | --- | --- | --- |
|  | Est | SE | Est | SE | Est | SE |
| *Fixed* | | | | | | |
| Intercept | 2.458 | (0.026) | 4.265 | (0.014) | -1.107 | (0.022) |
| Prior exclusion | -0.875 | (0.054) | 0.671 | (0.058) | 1.386 | (0.077) |
| *Random* | | | | | | |
| School Variance | 0.190 | (0.016) | 0.049 | (0.005) | 0.122 | (0.012) |
| Student Variance | 0.805 | (0.005) | 0.948 | (0.006) | 0.867 | . |
| *Statistics* | | | | | | |
| VPC | 19% | | 5% | | 12% | |
| R-Squared | 0.005 | | 0.003 | | 0.012 | |

Note. The results are standardised onto a common response scale (with mean 0 and SD 1) to facilitate comparisons across the three student outcomes.

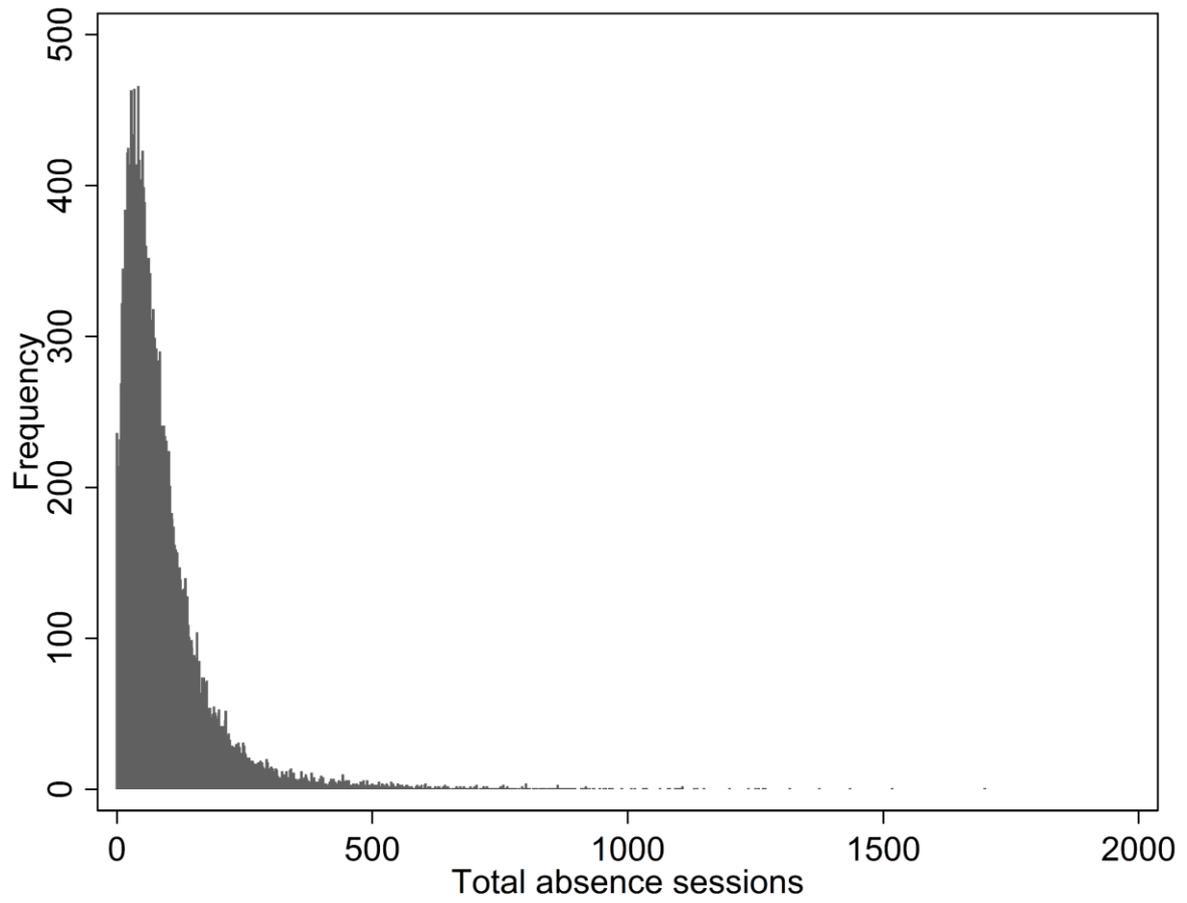

Supplementary Figure S1. Distribution of student total absence sessions over the five years of secondary schooling

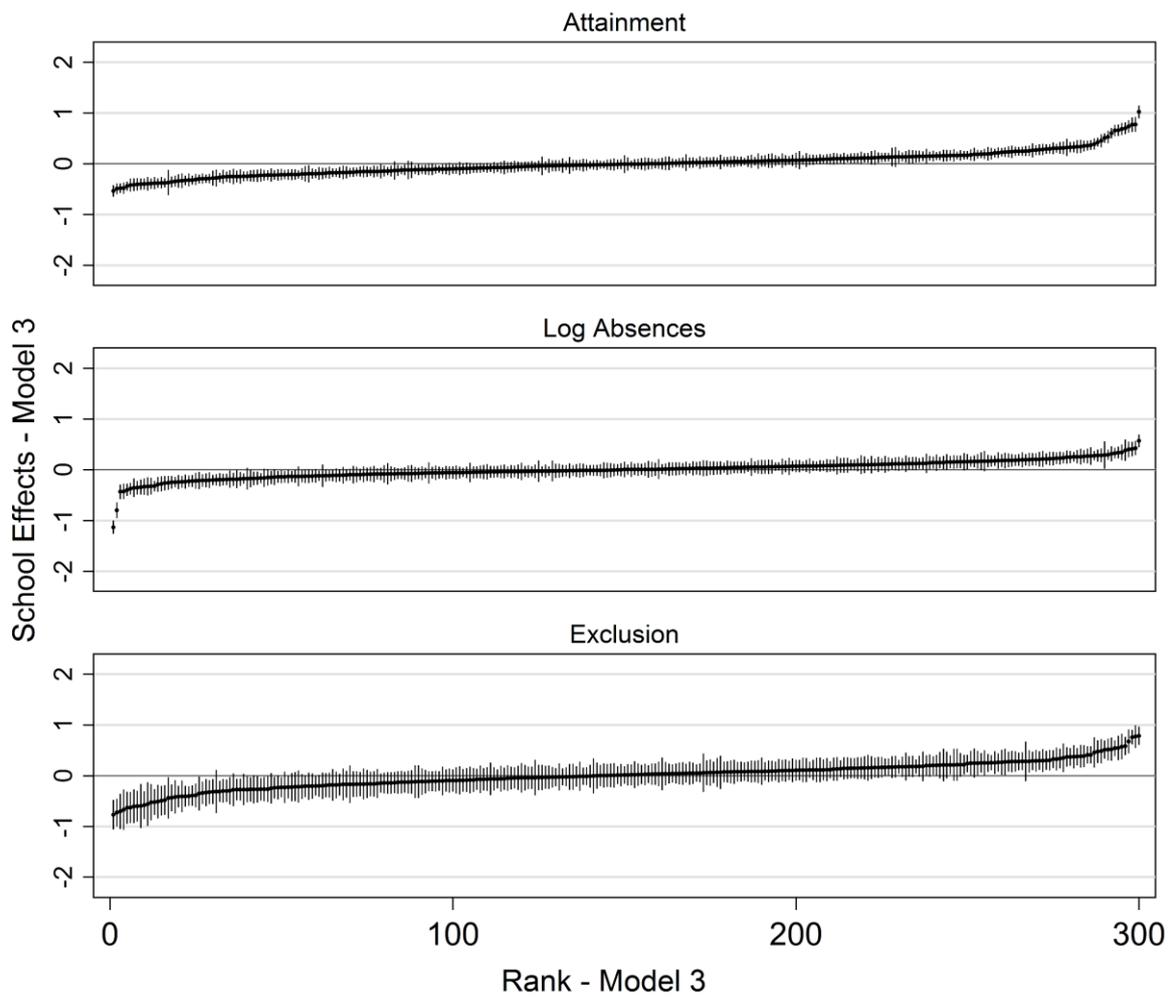

Supplementary Figure S2. Caterpillar plots for Model 3: contextual value-added model.

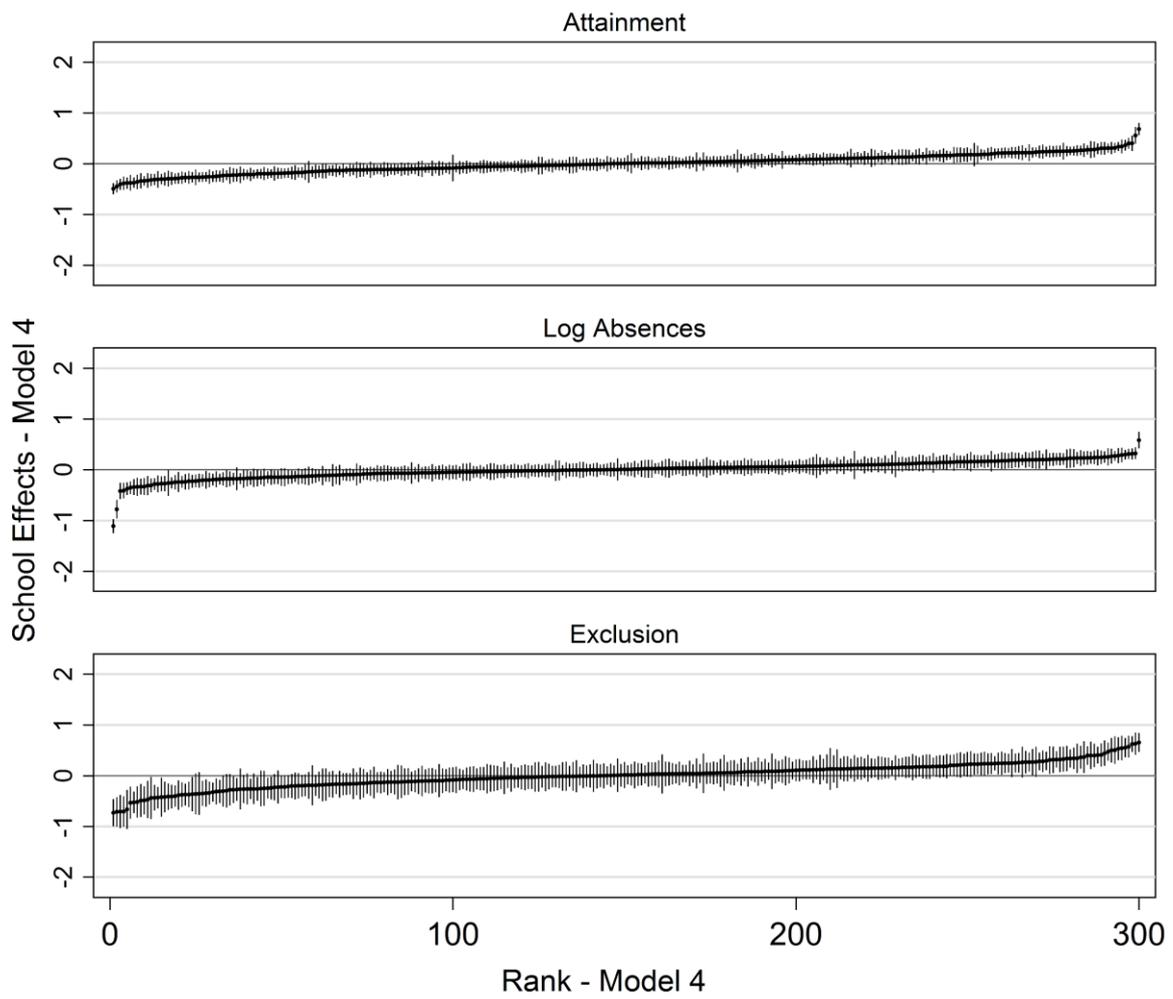

Supplementary Figure S3. Caterpillar plots for Model 4: contextual value-added model with school characteristics.

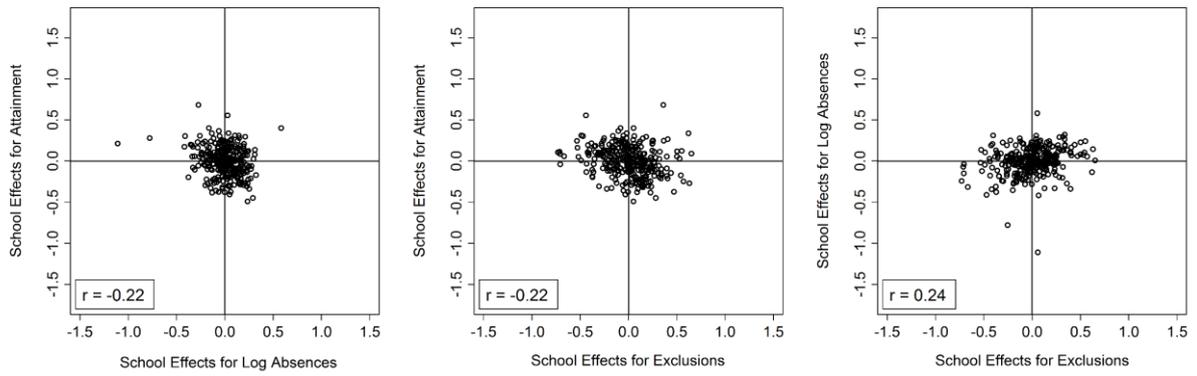
Supplementary Figure S4. Scatterplots of school effects between attainment, log absences and exclusions for Model 4: contextual value-added model with school characteristics.